\documentclass[twocolumn]{revtex4}

\usepackage{amsmath}    % need for subequations
\usepackage{graphicx}   % need for figures
\usepackage{verbatim}   % useful for program listings
\usepackage{color}      % use if color is used in text
\usepackage{subfigure}  % use for side-by-side figures
\def\lowsim{\mathrel{\lower 0.7 ex \hbox to 0 pt{$\sim$\hss}}}

 \newcommand{\wpi}{\omega_{pi}}

\begin{document}

\title{Reynolds Number and Intermittency in the Expanding Solar Wind:
Predictions Based on Voyager Observations}
\author{T. N. Parashar, M. Cuesta, W. H. Matthaeus}
\affiliation{Bartol Research Institute, Department of Physics and Astronomy, University of Delaware, Newark, DE}

\begin{abstract}
The large scale features of the solar wind are examined in order to
predict small scale features of turbulence in unexplored regions of the
heliosphere. The strategy is to examine how system size, or effective
Reynolds number, varies, and then how this quantity influences observable
statistical properties, including intermittency properties of solar wind
turbulence.  The expectation based on similar hydrodynamics scalings,
is that the kurtosis, of the small scale magnetic field increments, will
increase with increasing Reynolds number.  Simple theoretical arguments
as well as Voyager observations indicate that effective interplanetary
turbulence Reynolds number decreases with increasing heliocentric
distance. The decrease of scale-dependent magnetic increment kurtosis
with increasing heliocentric distance, is verified using a newly refined
Voyager magnetic field dataset. We argue that these scalings continue to
much smaller heliocentric distances approaching the Alfv\'en critical
region, motivating a prediction that the Parker Solar Probe spacecraft
will observe increased magnetic field intermittency, stronger current
sheets, and more localized dissipation, as its perihelion approaches the
critical regions. Similar arguments should be applicable to turbulence
in other expanding astrophysical plasmas.
\end{abstract}

\maketitle
\section{Introduction}

How can scaling of Voyager observations of interplanetary turbulence
inform predictions about plasma turbulence in a broader context?
The answer to this question is important not only to anticipate what
will be observed by spacecraft such as Parker Solar Probe as well as
Solar Orbiter, but also for observation of expanding plasmas in diverse
astrophysical situations.  We explore this possibility here, with specific
predictions for Parker Solar Probe\cite{FoxSSR16}, by employing Voyager
magnetic field observations\cite{MatthaeusJGR82,NessJGR01}.  The data
allow examination of the behavior of effective turbulence Reynolds
number and the kurtosis of magnetic field fluctuations over a wide range
of heliocentric distances.  We find that the Reynolds number decreases
and the kurtosis at a fixed physical scale decreases with increasing
distance, with associated expectations concerning the roughness of the
magnetic field. Extrapolating the observed scaling to lower heliocentric
distances motivates the prediction that the spatial concentration of
coherent structures increases approaching the Alfv\'en critical region
from outside, an effect that Parker Solar Probe should soon measure.
This letter provides theoretical and observation details that motivate
this prediction, and discusses further applications.

Theories of turbulence, in particular Kolmogorov theory
\cite{Kolmogorov41a, Kolmogorov41c, KolmogorovJFM62, FrischBook} and its
many variations, are frequently applied to understanding {\it in situ}
spacecraft observations in the interplanetary plasma\cite{BrunoLRSP13,
HorburyPRL08, ChenJPP16}.  In most instances these theories are invoked,
either explicitly or implicitly, in a form appropriate the regime
of {\it universality} conjectured to obtain in the limit of infinite
Reynolds number \cite{Kolmogorov41a, Kolmogorov41c, KolmogorovJFM62}.
However it is well known based on experimental data, especially in
hydrodynamics\cite{PopeBook}, that the dimensionless Reynolds number,
at attainable finite values, controls scaling of numerous statistical
quantities, including the approach to an asymptotic dissipation rate and
the scaling of higher order moments or increments.  There is substantial
evidence mainly based on simulations, that magnetohydrodynamics (MHD)
and other plasma models exhibit analogous systematic variations with
Reynolds-like numbers \cite{LinkmannPRL15, BandyopadhyayPRX18}.  Here we will examine turbulence in the solar wind,
quantifying variation of a specific intermittency parameter, the kurtosis,
as an effective Reynolds number is varied.

In hydrodynamics the value of kinematic viscosity $\nu$ enters
into the definition of Reynolds number $R=uL/\nu$, for turbulence
speed $u$, and energy containing (outer) scale $L$.  For the weakly
collisional plasma found in the solar wind, $\nu$ is meaningless,
but one may adapt the notion of Reynolds number by exploiting its
relationship to ``system size'', meaning the extent of the inertial
range.  In Kolmogorov theory, Reynolds number also relates the outer
(correlation) scale $L$ and inner (dissipation) scale $\eta$, as $Re
= (L/\eta)^{4/3}$.  For a weakly collisional plasma, the Kolmogorov
dissipation scale $\eta$ maybe reasonably replaced by the ion inertial
scale $d_i$ (or the thermal gyroradius if plasma beta becomes large),
given that the observed inertial range at MHD scales terminates at
the largest proton kinetic scale encountered by the direct energy
cascade \cite{LeamonJGR98,ChenGRL14}.  Accordingly we adopt a
definition of effective Reynolds number in terms of the system size
$L/d_i$, or size of the inertial range, as $Re=(L/d_i)^{4/3}$.

Below we will employ this definition of effective Reynolds number and
examine its behavior between 1 AU and 10 AU in Voyager magnetic field
data. The observational finding, backed by an elementary theoretical
assessment, is that $Re$ decreases with increasing heliocentric distance
in the solar wind.  Thus, even as the wind expands to fill the available 
volume, and in this sense becomes larger, the ``system size'' from the
perspective of turbulence, is decreasing.

Based on the determination of the radial behavior of Reynolds number,
the next step will be an assessment of the behavior of a normalized
fourth order moment of the magnetic fluctuations, and its scaling with the
Reynolds umber.  Kurtosis, the normalized fourth moment, measures the
roughness of the magnetic field and appearance of coherent structures.
The kurtosis of increments of a magnetic field cartesian component
time-series $b(t)$ with time lag $\tau$ is defined as $\kappa(\tau)
= <\Delta b(\tau)^4>/<\Delta b(\tau)^2>^2$ where $\Delta b(\tau) =
b(t+\tau)-b(t)$ is the increment at a scale $\tau$.  Following typical
practice in solar wind studies, we will exploit the highly supersonic
and super-Alfv\'enic flow at speed  $V_{sw}$ to interpret statistical
properties at time lags $\tau$ with spatial lags $r= -V_{sw} \tau$,
the so-called Taylor hypothesis.  The use of Taylor's hypothesis has
been shown to work well for first and second order statistics down to
the ion inertial scale and smaller\cite{ChhiberJGR18}.

{\it Some expectations based on theory and observations.} 
%\section{Dataand Methodology \label{data}} 
The expected variation of Reynolds number with heliocentric
distance may be anticipated using simple arguments as follows:
A von Karman-Howarth phenomenology has been shown to work well for
explaining radial variations of turbulence and plasma properties of
the solar wind, including correlation length and proton temperature
\cite{ZankJGR96,BreechJGR08}.  Temporarily ignoring expansion, a
relevant pair of equations\cite{MatthaeusJPP96} is $\frac{dZ^2}{dt}
= -\frac{1}{2} \frac{Z^3}{L}$ and $\frac{dL}{dt} = Z$, where $Z$ is
turbulence amplitude and $L$ the correlation scale. The solution behaves
as $L(t) \sim \sqrt{t}$, which when employing the Taylor hypothesis
becomes $L(R) \sim R^{1/2}$ where here $R$ is heliocentric distance.
Meanwhile to a reasonable approximation the proton number density in
the solar wind falls off as $n(R) \sim R^{-2}$, and by definition the
proton inertial scale in the expanding solar wind behaves as $d_i(n) \sim
n^{-1/2} \sim R$.  Therefore the Reynolds number may behave approximately
as $Re = (L/d_i)^{4/3} \sim (R^{1/2}/R)^{4/3} \sim R^{-2/3}$.

On the observational side, it is well established that the correlation
scale varies with heliocentric distance\cite{KleinSW92,ZankJGR96}.
This variation is found\cite{RuizSP14} to be approximately $\sim R^{0.44}$
in a mixed latitude ensemble, although the results did not have strong
dependence on plasma beta.

Turning to the kurtosis at small spatial lags, we find very little in
the MHD or plasma literature concerning expectations for its behavior as
Reynolds number is varied. There are studies at fixed (or, uncontrolled)
Reynolds numbers, of the scale dependent kurtosis of primitive variables
in simulations, and in solar wind and magnetosheath observations
\cite{SorrisoValvoGRL99, MacekApJL17, ChhiberJGR18}.  Some insight into
scaling of kurtosis at kinetic scales has been obtained in relatively
low Reynolds number kinetic simulations \cite{ParasharApJ15}.  There are
also studies of multi-fractal scalings, but again, as far as we are
aware, always without regard for Reynolds number or its variation.
In hydrodynamics the situation is more advanced both in experiments
and in theory.  The small scale kurtosis of longitudinal increments
should at zero separation approach the kurtosis of longitudinal
spatial {\it derivatives}. Below we will carry out an analysis of
increments in the smaller inertial range scales of the solar wind,
arguably small enough that the kurtosis approaches that of the magnetic
field derivatives.  In this regard, we take note of a collection of
well studied hydrodynamic experiments\cite{VanAttaPFL80}.  The best
fits of these data to the variation of $\kappa \sim Re^{\gamma}$ is
$\gamma \approx 0.16$ to $0.2$. This nicely brackets an analytical
estimate, based on phenomenological estimates as well as extensions
of Kolmogorov-Obukhov treatments based on log-normal distributions
of increments \cite{OboukhovJFM62}.  These theoretical estimates are
discussed in the same reference\cite{VanAttaPFL80} and lead to, for
example, $\gamma = 3 \mu/4 = 3/16$ when the log-normal intermittency
parameter $\mu$ takes the value 0.25.  These hydrodynamic expectations
anticipate the similar behavior we find below for solar wind turbulence.

{\it Voyager Data.}
The Voyager magnetic field datasets are an excellent choice for the
present study due the wide range of distances spanned: the associated
variation of correlation scale and density are found to give a systematic
variation of Reynolds number.  Therefore we use 1.92s cadence magnetic
field vector data from the Voyager 1 spacecraft. For the present study,
we use a refined dataset, explained below, covering the range from 1 AU
up to $\sim 10 AU$.

The publicly available Voyager magnetic field dataset, at the time of this
writing, requires additional work to make it useful for a statistical
study.  After the cleanup procedure we employed this improved subset of
Voyager 1 MAG data will be made available for other purposes.

One problem with the previously available data is due to the large data
gaps associated with regular lack of telemetry.  The first task we perform
is to make the time series uniform by filling in the missing data points
with NaNs (not-a-numbers). The final time series has $\sim 60$ million 
data points.

The data available from the NASA-NSSDC also have many intervals containing
``calibration rolls'' as well as unexplained ``noise'' that need to
be removed before any reliable turbulence statistics can be computed.
We clean the data using a series of techniques.  These include
application of obvious cutoffs (-50nT,50nT), a Hampel filter where the
outliers are replaced with NaNs instead of a median value, and finally,
visual inspection of high kurtosis regions to identify and filter out
remaining bad data.  We describe the technique in detail in a longer paper
\cite{CuestaApJ19}. The final clean time series has $\sim 23.27$ million
data points and $\sim 37.74$ million NaNs representing missing data
or discarded bad data.

After the cleaning procedure, we bin the data into 450 bins, each of
size 0.02AU. A significant number of these bins have no physical data
points, and some have very few physical data points. For statistical
significance, we exclude bins with less than 10000 physical data
points from our analysis. The remaining useful dataset 
comprises 328 bins, each having $>$10000 points. 
For data quality purposes, we then compute the kurtosis at
a small lag for the data in each bin. %
%Kurtosis of a magnetic component $B$
%is defined as $\kappa(\tau) = <\delta B(\tau)^4>/<\delta B(\tau)^2>^2$,
%where $\delta B(\tau)=B(t+\tau)-B(t)$ is the increment at lag $\tau$. 
This helps further identify bins with potential bad data points as $\kappa$
would be anomalously high for large unphysical discontinuities.  
Some retained bins contain what appears to be 
upstream waves \cite{SmithJGR83} or similar plasma activity that cause 
$\kappa$ to attain anomalously small values. 
We recall that 
using single spacecraft data, we employ increments at very small
time lags, and the Taylor hypothesis, to estimate radial magnetic
field derivatives.

{\it Analysis procedure: Reynolds number.}
The two-time vector autocorrelation function of magnetic field is defined as 
\begin{equation}
   C_R(\tau) =
   \frac{<\mathbf{b}(t)\cdot\mathbf{b}(t+\tau)>}{<|\mathbf{b}(t)|^2>},
\end{equation} 

where ${\bf b}= {\bf B} - \langle {\bf B} \rangle$.
The outer scale is computed, using the Taylor frozen-in flow hypothesis,
as the scale where the autocorrelation drops to $1/e$.  In particular,
we estimate the correlation scale as $L = V_{sw}\tau_o$ where $\tau_o$
is the lag at which the two-time correlation of the magnetic field
decreases to $1/e$.  With these definitions, the effective Reynolds
number can be computed as $Re = \left(\frac{L}{d_i}\right)^{4/3}$
where $d_i$ is the ion inertial length $c/\wpi$, chosen to represent
the scale at which the inertial range terminates.  The ion (proton)
inertial length is computed in each bin by appropriate averages of the
number density recorded by the Voyager plasma (PLS) plasma instrument.
Note that choosing the electron inertial length as the inner scale would
simply shift the Reynolds number by a constant value.

{\it Results:} 
\begin{figure}[!hb]
\includegraphics[width=\columnwidth]{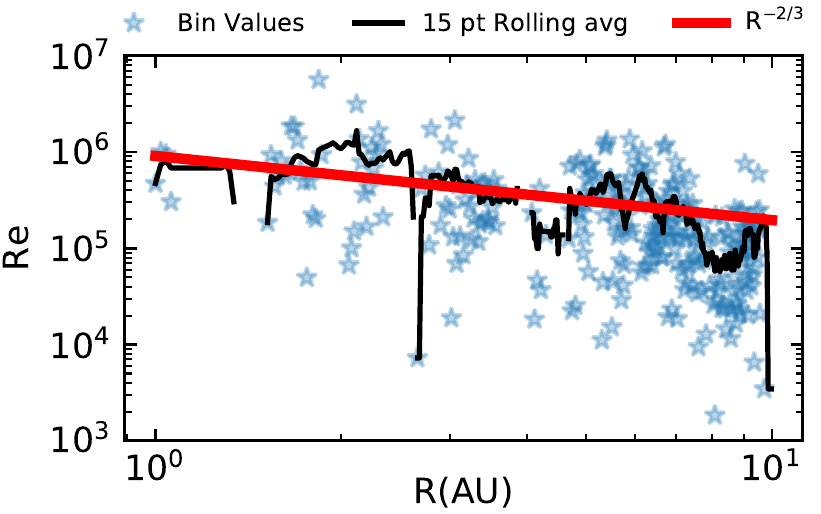}
\caption{(Color online) Effective Reynolds number as a function of heliocentric
distance. Blue stars represent value within a local bin. Black curve
is a 15 point running average of these values. The red line indicates
the anticipated decrease based on the elementary estimate presented in
the text.}
\label{Re-vs-r}
\end{figure}

Figure \ref{Re-vs-r} shows the effective Reynolds number computed
from the magnetic field, as a function of heliocentric distance in
the 1-10 AU Voyager 1 dataset. Blue stars represent the local value
of Reynolds number in each of the 450 bins, each  
approximately 0.02 AU wide in heliocentric distance.  Significant
variability is present in the Reynolds number, consistent with variable
turbulence conditions\cite{TuSSR95}, such as stream structure, coronal
mass ejections, etc.  There are some bins in which wave activity may
modify the results. (See Supplementary Material.)  To get a better view
of the average behavior of $Re$, we plot a 15 point running average
of the blue stars. A clear trend for $Re$ to drop with heliocentric
distance can be seen.  %except for the unusual drops near the planets.
The elementary estimate of $Re \sim R^{-2/3}$ based on $L\sim\sqrt{R}$
and $d_i \sim R$, is over-plotted as a thick red line. We emphasize
the point that {\em this is not a fit of any kind}: the average $Re$ follows
a trend similar to the simple theoretical prediction.  The turbulence
``system size'' or effective Reynolds number in the expanding solar wind
in fact decrease with increasing radial distance.

According to the hydrodynamic analogy, a systematic decrease in
$Re$ should be reflected in a decrease in the kurtosis of increments
computed at very small lags.  This expectation would be obtained whenever
the turbulence is fully developed, i.e., it is not too close to its
injection or initiation.  Having observed a systematic decrease in $Re$,
we now examine the small scale magnetic kurtosis in the same dataset as
a surrogate for the kurtosis of the longitudinal magnetic derivative.
We compute the kurtosis at a lag of $10 d_i$, where the distance is
computed from time lag, using Taylor's hypothesis, using the average
radial plasma velocity for the conversion.

\begin{figure}[!hb]
\includegraphics[width=\columnwidth]{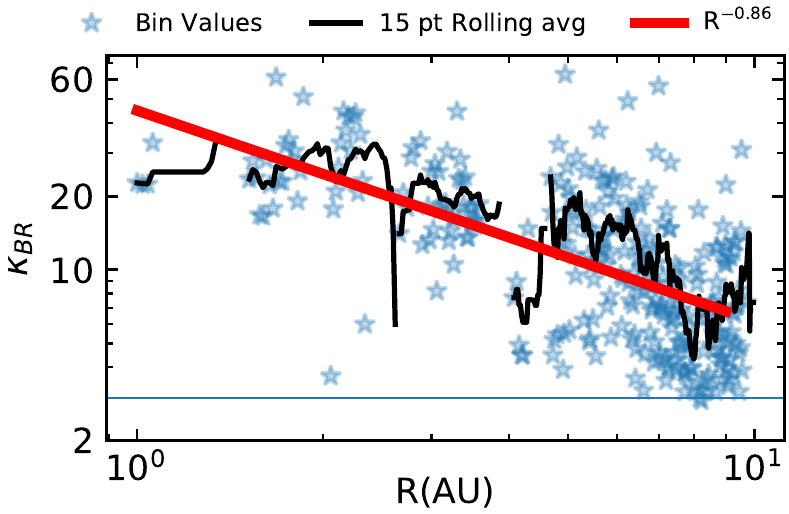}
\caption{(Color online) Kurtosis of $B_R$ at a lag of 10$d_i$. Blue stars represent
value inside a local bin, black curve is a 15 point running average of
these values. Horizontal blue line represents the kurtosis for Gaussian
noise. Red line is a power-law fit to the points as there is yet no
theory for how kurtosis should vary with Reynolds number in a plasma.}
\label{K-10di-vs-r}
\end{figure}

Figure \ref{K-10di-vs-r} shows the kurtosis of the increment of the
radial magnetic field component at the scale 10$d_i$ as a function of
heliocentric distance. Blue stars show value of kurtosis in each of the
450 bins, the black curve shows a 15 point running average, and the red
line shows a power-law fit to the data with an approximate dependence
$\kappa \sim R^{-0.86}$.  As expected, the kurtosis drops for larger
heliocentric distance, along with  the decrease of effective Reynolds
number with increasing distance.  Points that have lower kurtosis appear
in the same regions where $Re$ drops.  However these anomalously low
points do not significantly affect the overall conclusion that both
$Re$ and $\kappa$ decrease systematically with increasing heliocentric
distance.

{\it Conclusions:}
The above study, based on analysis of Voyager 1 magnetic field data,
shows two main results that are established firmly by the observations.
First, (I) the effective Reynolds number, as defined based on the extent
of the inertial range, decreases systematically in the interplanetary
medium between 1AU and 10 AU.  Second (II) the kurtosis of magnetic
component increments defined over the same range is also found to decrease
with increasing heliocentric distances.

Potentially interesting physics comes in examining the relationship
between these results. In fact, the reasoning that connects these two
findings may be cast in more than one framework:

First, one might simply assume that the effective Reynolds number is fully
equivalent to the ordinary Reynolds number, and then further assert that
a relationship such as the hydrodynamic relation $\kappa \sim Re^\gamma
\sim Re^{2\mu/4}$ also obtains for a weakly collisional plasma. As we
pointed out earlier, such a scaling can be obtained on purely empirical
grounds. Or, it can be deduced from Refined Similarity\cite{KolmogorovJFM62},
augmented by Obukhov's scaling hypothesis\cite{OboukhovJFM62} that
parameterizes anomalous scaling of increments with $L/\eta$ (which
here becomes $L/d_i$ by a separate argument.)  While this reasoning
may appeal more strongly to formal turbulence theory, we recall that
a Refined Similarity hypothesis has not been firmly established for a
collisionless plasma (although there have been preliminary discussions
of this\cite{MerrifieldPP05,ChandranApJ15}).  So the linkage in this
chain of reasoning, while appealing, is a bit tenuous.

A second argument for the relationship of results I and II rests on
understanding how the nonlinearities in turbulence, the most essential
of which are {\it quadratic}, give rise dynamically to the formation
of coherent structures.  Nonlinear spectral transfer forms coherent
structures, without direct involvement of dissipation, and progressively
at smaller scale \cite{FrischJMTAS83,WanPP09,WanPRL12}.  This is
evidenced, in both ideal and dissipative MHD, by the monotonic increase
in {\it filtered} kurtosis (high pass filtered) at scale $\ell$, as
the band-pass scale decreases.  Given that transfer is mainly local
in scale\cite{VermaPP05}, it follows that an inertial range with
greater bandwidth will incorporate a greater number of octaves of transfer
over which the coherent structures may form and intensify. Thus, larger
systems, i.e., larger $L/\eta$ or $L/d_i$, will have stronger coherent
structures and higher small scale kurtosis.  But this implies immediately
that $\kappa(10d_i)$ will decrease with effective Reynolds number as we
have defined it.  It is noteworthy that this second line of reasoning
takes no explicit position on the relationship between coherent structures
and dissipation, nor does it assume a refined similarity hypothesis,
in contrast to the first line of reasoning.  For  this reason we prefer,
at this time, the second track for connecting results I and II, although
we do not doubt that the more formal relationships may be established
more firmly in the future.

Having completed this excursion into the turbulence theoretic basis
for relating I and II, we now may view the behavior of the kurtosis
at 10$d_i$ as a {\it consequence} of the decrease in effective $Re$ at
larger heliocentric radial distance.  Such a connection has interesting
implications beyond magnetic field observed by Voyager in the outer heliosphere.

One major implication is the potential for extrapolating these results
to other expanding astrophysical plasmas.  We would expect, based on the
arguments above that other systems engaged in an approximate spherical
expansion, with evolving von Karman turbulence, will also
admit a baseline estimate of effective Reynolds number scaling as $Re
\sim R^{-2/3}$.  Turbulence properties that scale with Reynolds number
will systematically respond to this scaling, enabling in principle,
a variety of predictions for expanding systems such as galactic winds,
supernova remnants, etc.  
The decrease in kurtosis, or inverse filling factor,
of coherent current structures, is only one such prediction.

Closer to home, we are tempted to extrapolate the present results {\it
inward}, toward the corona, but outside the Alfv\'en critical region.
Such an extrapolation finds some partial support in correlation length
scalings in radius from Helios and Ulysses \cite{RuizSP14} that behave
as $L \sim R^{0.43}$ and density scalings \cite{McComasJGR00} that
remains close to $n\sim R^{-2}$ everywhere in the heliosphere on average.
Encouraged by this, one may extrapolate that effective Reynolds number
{\it increases} moving towards the inner interplanetary region where
the magnetic control imposed by the corona gives way to the turbulent
solar wind \cite{DeForestApJ16}.  Approaching this region from outside,
we expect that as $Re$ increases, the kurtosis at small scales (multiples
of $d_i$) will {\it increase}.  This becomes, in effect a prediction
for the recently launched Parker Solar Probe and the upcoming Solar
Orbiter spacecraft: as these missions explore the heliosphere approaching
the Alfv\'en critical region, we expect an increase in the frequency
and intensity of magnetic discontinuities and current sheets. This
should be indicted by higher order statistics such as the kurtosis at
small scales. There may be associated implications such as stronger
concentrations of dissipation and heating \cite{OsmanPRL12a},
a possibility we have not pursued here. 
There are reasons to believe that the turbulence is sustained well 
below the the Alfv\'en critical surfce \cite{AdhikariApJ19} but the
nature of turbulence and its driving like differs in that region (the corona), and we do not attempt to extend our prediction
beyond this point at present.
As a step toward supporting the present
prediction, we are currently processing Helios data to confirm this effect
down to 0.3AU. Results will be presented in a separate publication,
as we anxiously wait the relevant Parker Solar Probe analyses.  We may
anticipate that there will be further interesting consequences of the
systematic variation of interplanetary effective Reynolds number that
will he examined in future study.

\begin{acknowledgments} The authors are thankful to Chuck Smith for useful discussions. 
This research supported in part by the
NASA Heliophysics Guest Investigator Program (80NSSC19K0284,  NNX17AB79G) and Supporting Research (80NSSC18K1648) and by the ISOIS
Parker Solar Probe Project though Princeton subcontract SUB0000165.
\end{acknowledgments}

%\bibliography{parashar-ref,tulasi-cpubs,tulasi-jpubs,tulasi-mpubs,tulasi-upubs}

\end{document}